%% file: lower_upper_conditioning.tex
\documentclass{eptcs}

\newif\ifignore 
\ignorefalse
\newcommand{\auxproof}[1]{
\ifignore\mbox{}\newline
\textbf{PROOF:} \dotfill\newline
{\it #1}\mbox{}\newline
\textbf{ENDPROOF}\dotfill
\fi}

\usepackage{breakurl}
\usepackage{etex} 
\usepackage{amsthm}
\usepackage{amsmath}
\usepackage{amssymb}
\usepackage{amstext}
\usepackage{mathtools}
\usepackage{mathrsfs}
\usepackage{stmaryrd}
\usepackage{xspace}
\usepackage{enumitem}
\usepackage{cmll}
\usepackage{hyperref}
\usepackage{url}
\usepackage{nicefrac}
\usepackage{fancybox}
\usepackage{listings} 
\usepackage{wrapfig}
\usepackage{xypic}
\usepackage[all,cmtip]{xy}
\xyoption{2cell}
\UseTwocells
\xyoption{tips}
\usepackage{tikz}
\usetikzlibrary{circuits.ee.IEC}
\usetikzlibrary{shapes,shapes.geometric}

\input{tikzdefs}

\newtheorem{rem}{Remark}

\newcommand{\mystretch}{%
  \def\arraystretch{1.3}%
  \setlength{\arraycolsep}{2pt}%
}

\newcommand{\QEDbox}{\square}
\newcommand{\QED}{\hspace*{\fill}$\QEDbox$}

\makeatletter
\newcommand{\mathoverlap}[2]{\mathpalette\mathoverlap@{{#1}{#2}}}
\newcommand{\mathoverlap@}[2]{\mathoverlap@@{#1}#2}
\newcommand{\mathoverlap@@}[3]{\ooalign{$\m@th#1#2$\crcr\hidewidth$\m@th#1#3$\hidewidth}}
\makeatother

\newcommand{\klafter}{\mathrel{\bullet}}

\DeclareSymbolFont{T1op}{T1}{cmr}{m}{n}
\SetSymbolFont{T1op}{bold}{T1}{cmr}{bx}{n}
\DeclareMathSymbol{\mathguilsinglleft}{\mathopen}{T1op}{'016}
\DeclareMathSymbol{\mathguilsinglright}{\mathclose}{T1op}{'017}

\newcommand{\idmap}[1][]{\ensuremath{\mathrm{id}_{#1}}}
\newcommand{\after}{\mathrel{\circ}}
\newcommand{\evmap}{\mathrm{ev}}

\newcommand{\tr}{\ensuremath{\mathrm{tr}}}

\newcommand{\asrt}{\ensuremath{\mathrm{asrt}}}

\newcommand{\pair}{\ensuremath{\mathrm{pair}}}
\newcommand{\proj}{\ensuremath{\mathrm{proj}}}
\newcommand{\extract}{\ensuremath{\mathrm{extr}}}

\newcommand{\setin}[3]{\{#1\in#2\;|\;#3\}}
\newcommand{\supp}{\mathrm{supp}}

\newcommand{\all}[2]{\forall#1.\,#2}

\newcommand{\ex}[2]{\exists#1.\,#2}

\newcommand{\tuple}[1]{\langle#1\rangle}

\newcommand{\ket}[1]{\ensuremath{|{\kern.1em}#1{\kern.1em}\rangle}}
\newcommand{\bigket}[1]{\ensuremath{\big|{\kern.1em}#1{\kern.1em}\big\rangle}}
\newcommand{\bra}[1]{\langle\,#1\,|}
\newcommand{\one}{\ensuremath{\mathbf{1}}}

\newcommand{\andthen}{\mathrel{\&}}

\newcommand{\C}{\mathbb{C}}

\newcommand{\marg}[2]{\ensuremath{{#1}\mathrel{\%}{#2}}}
\newcommand{\margsign}{\mathsf{M}}

\newcommand{\HSip}[2]{\ensuremath{\tuple{#1|#2}_{\mathrm{HS}}}}

\newcommand{\distributionsymbol}{\mathcal{D}}

\newcommand{\Dst}{\distributionsymbol}

\newcommand{\UF}{\ensuremath{\mathcal{U}{\kern-.75ex}\mathcal{F}}}

\newcommand{\Cat}[1]{\ensuremath{\mathbf{#1}}\xspace}

\newcommand{\Kl}{\mathcal{K}{\kern-.4ex}\ell}
\newcommand{\EM}{\mathcal{E}{\kern-.4ex}\mathcal{M}}

\newcommand{\Sets}{\Cat{Sets}}

\newcommand{\Ef}{\ensuremath{\mathcal{E}{\kern-.5ex}f}}
\newcommand{\intd}{{\kern.2em}\mathrm{d}{\kern.03em}}

\newcommand{\OF}{\ensuremath{\mathcal{O}{\kern-.1em}\mathcal{F}}}
\newcommand{\Closed}{\ensuremath{\mathcal{C}{\kern-.05em}\ell}}

\makeatother
 \newdir{ >}{{}*!/-7.5pt/@{>}}
 \newdir{|>}{!/4.5pt/@{|}*:(1,-.2)@^{>}*:(1,+.2)@_{>}}
 \newdir{ |>}{{}*!/-3pt/@{|}*!/-7.5pt/:(1,-.2)@^{>}*!/-7.5pt/:(1,+.2)@_{>}}

\newsavebox\sbground
\savebox\sbground{\begin{tikzpicture}[circuit ee IEC,yscale=0.5,xscale=0.4]
                \draw (0,-2ex) to (0,0) node[ground,rotate=90,xshift=.65ex] {};
                \end{tikzpicture}}
\newcommand\ground{\mathbin{\text{\raisebox{-0.2ex}{\usebox\sbground}}}}

\newsavebox\sbuniform
\savebox\sbuniform{\begin{tikzpicture}[circuit ee IEC,yscale=0.5,xscale=0.4]
                \draw (0,2ex) to (0,0) node[ground,rotate=-90,xshift=.65ex] {};
                \end{tikzpicture}}
\newcommand\uniform{\mathbin{\text{\raisebox{-0.2ex}{\usebox\sbuniform}}}}

\newsavebox\sbcopier
\savebox\sbcopier{%
  \begin{tikzpicture}[baseline=0pt]
    \node[copier,scale=.7] (a) at (0,3.6pt) {};
    \draw (a) -- +(-90:.16);
    \draw (a) -- +(45:.19);
    \draw (a) -- +(135:.19);
  \end{tikzpicture}}
\newcommand{\copier}{\mathord{\usebox\sbcopier}}

\newcommand{\eg}{\textit{e.g.}\xspace}

\newcommand{\derivationsign}[1]{\ensuremath{\mathrel{\vdash_{\kern-.4em\scriptscriptstyle\mathsf{#1}}}}}

\newcommand{\Prob}{\mathop{\mathsf{P}}}

\newcommand{\inhab}{\ensuremath{\kern-.2em:\kern-.2em}}
\newcommand{\tsum}{\ensuremath{\kern-.2em+\kern-.2em}}

\newcommand{\pklar}{\ar@{-^>}|-@{|}}

\DeclareFixedFont{\ttb}{T1}{txtt}{bx}{n}{12} 
\DeclareFixedFont{\ttm}{T1}{txtt}{m}{n}{12}  
\usepackage{color}
\definecolor{deepblue}{rgb}{0,0,0.5}
\definecolor{deepred}{rgb}{0.6,0,0}
\definecolor{deepgreen}{rgb}{0,0.5,0}

\newcommand\pythonstyle{\lstset{
backgroundcolor = \color{lightgray},
language=Python,
basicstyle=\ttm,
otherkeywords={self,>>>},             
keywordstyle=\ttb\small\color{deepblue},
emph={@,+\%%
literate={.+}{{{\color{red}.+}}}2 {.**}{{{\color{red}.\**{}}}}2 {*}{{{\color{red}*}}}1
},          
emphstyle=\ttb\small\color{deepred},    
stringstyle=\small\color{deepgreen},
frame=tb,                         
showstringspaces=false            %
}}
\lstnewenvironment{python}[1][] 
{
\pythonstyle
\lstset{basicstyle=\normalfont\ttfamily\small#1}
}
{}

\newcommand\pythoninline[1]{\pythonstyle\lstinline[basicstyle=\normalfont\ttfamily\small]{#1}} 

\theoremstyle{plain}
\newtheorem{theorem}{Theorem}
\newtheorem{lemma}{Lemma}
\newtheorem{proposition}[lemma]{Proposition}

\theoremstyle{definition}
\newtheorem{definition}{Definition}

\newenvironment{myproof}[1][Proof]%
   { \begin{trivlist}%
     \item[\hskip \labelsep {\bfseries #1}]%
   }%
   { \end{trivlist}%
   }

\providecommand{\event}{QPL 2018}
\title{Lower and Upper Conditioning \\ in Quantum Bayesian
  Theory\thanks{The research leading to these results has received
    funding from the European Research Council under the European
    Union's Seventh Framework Programme (FP7/2007-2013) / ERC grant
    agreement nr.~320571.}}  
\author{Bart Jacobs \institute{Institute
    for Computing and Information Sciences (iCIS), \\ Radboud
    University Nijmegen, The Netherlands.}  \email{bart@cs.ru.nl} }

\def\titlerunning{Lower and Upper Conditioning}
\def\authorrunning{B. Jacobs}

\date{\small \today}

\begin{document}
\maketitle

\begin{abstract}
Updating a probability distribution in the light of new evidence is a
very basic operation in Bayesian probability theory. It is also known
as state revision or simply as conditioning. This paper recalls how
locally updating a joint state can equivalently be described via
inference using the channel extracted from the state (via
disintegration). 

This paper also investigates the quantum analogues of conditioning,
and in particular the analogues of this equivalence between updating a
joint state and inference. The main finding is that in order to
obtain a similar equivalence, we have to distinguish two forms of
quantum conditioning, which we call lower and upper conditioning. They
are known from the literature, but the common framework in which we
describe them and the equivalence result are new.
\end{abstract}

\section{Introduction}\label{sec:intro}

This paper is about quantum analogues of Bayesian reasoning.  It works
towards one main result, Theorem~\ref{thm:quantuminference} below,
which gives a relation between locally updating a joint state and
Bayesian inference. This is a fundamental matter, which requires some
preparation in order to set the scene.

We use the term `classical' probability for the ordinary, non-quantum
form. We often use the word `state' for a probability distribution,
both in the classical and the quantum case. Classical Bayesian
probability is based on what is called Bayes' rule. It describes
probabilities of events (evidence) in an updated state. In fact, there
are two closely related rules, sometimes called `product rule' en
`Bayes rule' (proper). Making this distinction is not so relevant in
the classical case, but, as we shall see, it is very relevant in the
quantum case.

The paper starts with the back-and-forth constructions between a joint
state (distribution) on the one hand, and a channel with an initial
state on the other. A channel is a categorical abstraction of a
conditional probability. We shall describe this process in terms of
pairing and disintegration, following~\cite{ChoJ17a}. This process has
a logical dimension that relates locally updating a joint state
(`crossover inference') and Bayesian inference via the associated
channel, in a result called the Bayesian Inference Theorem (see
Theorem~\ref{thm:classicalconditioning} below). This result is already
described in~\cite{ChoJ17a}, but is repeated here in more concrete
form, and illustrated with an example.

The second part of the paper is about analogues in the quantum world.
The constructions back-and-forth between a joint state and a channel
exist in the literature~\cite{LeiferS13} and are adapted to the
current context. What is new here is the quantum logical analogue of
this back-and-forth process. It is shown that updating a state with
new evidence, in the form of a predicate, splits in two operations,
which we call `lower' and `upper' conditioning. Both forms exist
already, but not as counterparts. We show that the earlier mentioned
product rule holds for lower conditioning, but Bayes' rule itself
holds for upper conditioning. In classical probability, the `lower'
and `upper' versions coincide.

In a next step, the main result of the paper
(Theorem~\ref{thm:quantuminference}) shows how `lower' updating a
joint state can equivalently be done via Bayesian inference with
`upper' conditioning, using the channel that is extracted from the
joint state. This puts lower and upper conditioning into perspective
and unveils some fundamental aspects of a quantum Bayesian theory.

Finally, there are two separate points worth emphasising. First, several
constructions in this paper are illustrated with concrete
calculations, via the Python-based tool EfProb~\cite{ChoJ17b}; it
works both for classical and quantum probability and uses a common
language for both. Next, along the way we find a novel result about
how disintegration introduces `semi' higher order structure in
discrete probability, see Subsection~\ref{subsec:excursion}.

\section{Basics of discrete classical probability}\label{sec:discrete}

This section recalls the basics of (classical, finite) discrete
probability and fixes notation. For more information we refer
to~\cite{Jacobs17a}. A \emph{distribution}, also called a
\emph{state}, on a set $X$ is a function $\omega\colon X \rightarrow
     [0,1]$ with finite support $\supp(\omega) =
     \setin{x}{X}{\omega(x) \neq 0}$ and with $\sum_{x}\omega(x) =
     1$. Such a distribution can also be written as formal convex sum
     $\omega = \sum_{x} \omega(x)\ket{x}$. We write $\Dst(X)$ for the
     set of such distributions. The mapping $X\mapsto\Dst(X)$ is a
     monad on the category of sets, called the distribution monad.

A \emph{joint} state is a state on an $n$-ary product set. A binary
state is thus a distribution $\tau\in\Dst(X_{1} \times X_{2})$. It has
first and second \emph{marginals}, written here as
$\margsign_{1}(\tau) \in \Dst(X_{1})$ and $\margsign_{2}(\tau)
\in\Dst(X_{2})$. These marginal states are defined in the standard way
as $\margsign_{1}(\tau)(x_{1}) = \sum_{x_{2}}\tau(x_{1}, x_{2})$ and
$\margsign_{2}(\tau)(x_{2}) = \sum_{x_{1}}\tau(x_{1}, x_{2})$. In the
other direction, two states $\omega_{i}\in\Dst(X_{i})$ can be combined
to product state $\omega_{1}\otimes\omega_{2}\in\Dst(X_{1}\times
X_{2})$ via $(\omega_{1}\otimes\omega_{2})(x_{1},x_{2}) =
\omega_{1}(x_{1}) \cdot \omega_{2}(x_{2})$. Obviously,
$\margsign_{i}(\omega_{1}\otimes\omega_{2}) = \omega_{i}$.

A \emph{channel} is a function of the form $c\colon X \rightarrow
\Dst(Y)$, that is, a map $X\rightarrow Y$ in the Kleisli category
$\Kl(\Dst)$ of the distribution monad $\Dst$. Such a channel $c$ has a
\emph{Kleisli extension} function, or \emph{state transformer}, $c \gg
(-) \colon \Dst(X) \rightarrow \Dst(Y)$ given by $(c \gg \omega)(y) =
\sum_{x} \omega(x) \cdot c(x)(y)$. For another channel $d\colon Y
\rightarrow \Dst(Z)$ there is a composite channel $d \klafter c \colon
X \rightarrow \Dst(Z)$ given by $(d \klafter c)(x) = d \gg
c(x)$. Channels $c_{i} \colon X_{i} \rightarrow \Dst(Y_{i})$ can be
combined to a product channel $c_{1}\otimes c_{2} \colon X_{1}\times
X_{2} \rightarrow \Dst(Y_{1}\times Y_{2})$ by $(c_{1}\otimes
c_{2})(x_{1}, x_{2}) = c_{1}(x_{1})\otimes c_{2}(x_{2})$.

A (fuzzy) predicate on a set $X$ is a function $p\colon X \rightarrow
[0,1]$.  For another predicate $q\in [0,1]^{X}$ there is a (sequential)
conjunction predicate $p\andthen q$ on $X$ via $(p\andthen q)(x) =
p(x)\cdot q(x)$. For two predicates $p_{i} \in [0,1]^{X_i}$ on
different sets $X_{i}$ we can form a parallel conjunction predicate
$p_{1}\otimes p_{2} \in [0,1]^{X_{1}\times X_{2}}$, given by
$(p_{1}\otimes p_{2})(x_{1},x_{2}) = p_{1}(x_{1}) \cdot
p_{2}(x_{2})$. There is always a truth channel $\one \in [0,1]^{X}$
given by $\one(x) = 1$.

For a state $\omega\in\Dst(X)$ and a predicate $p\in [0,1]^{X}$ on the
same set $X$ the validity $\omega\models p$ in $[0,1]$ is the expected
value $\sum_{x} \omega(x) \cdot p(x)$. If this validity is non-zero,
one can form a conditioned state $\omega|_{p}$ on $X$, given by
$\omega|_{p}(x) = \frac{\omega(x)\cdot p(x)}{\omega\models p}$. This
updated state $\omega|_{p}$ is called `$\omega$ given $p$', and is
commonly written as $\omega(-|p)$.  It is easy to check to see that
conditioning with truth does nothing: $\omega|_{\one} = \omega$.

\begin{proposition}
\label{prop:classicalbayes}
Assuming the conditionings of the states below exist, we have the
`product' rule on the left, and the `Bayesian' rule on the right:
\begin{equation}
\label{eqn:classicalbayes}
\begin{array}{rclcrcl}
\omega|_{p} \models q
& = &
\displaystyle\frac{\omega\models p\andthen q}{\omega\models p}
& \hspace*{5em} &
\omega|_{p} \models q
& = &
\displaystyle\frac{(\omega|_{q}\models p)\cdot(\omega\models q)}{\omega\models p}.
\end{array}
\end{equation}

\noindent Moreover, successive conditioning can be reduced to a
single conditioning, as on the left below, so that conditioning becomes
commutative, as on the right:
\begin{equation}
\label{eqn:classicalsuccesiveconditioning}
\begin{array}{rclcrcl}
(\omega|_{p})|_{q}
& = &
\omega|_{p\andthen q}
& \hspace*{5em} &
(\omega|_{p})|_{q}
& = &
(\omega|_{q})|_{p}.
\end{array}
\end{equation}
\end{proposition}

\begin{myproof}
The first equation in~\eqref{eqn:classicalbayes} follows simply by
unravelling the definitions. The second equation directly follows from
the first one, using that conjunction $\andthen$ is
commutative. Similarly one obtains the equations
in~\eqref{eqn:classicalsuccesiveconditioning}. \QED
\end{myproof}

Each channel $c \colon X \rightarrow \Dst(Y)$ also gives rise to a
\emph{predicate transformer} function $c \ll (-) \colon [0,1]^{Y}
\rightarrow [0,1]^{X}$, given by $(c \ll q)(x) = \sum_{y} c(x)(y)\cdot
q(y)$. We can now relate validity $\models$ and state/predicate
transformation ($\gg$ and $\ll$) via the following fundamental
equality of validities:
\begin{equation}
\label{eqn:classicalvaliditytransformation}
\begin{array}{rcl}
(c \gg \omega) \models q
& \;=\; &
\omega \models (c \ll q).
\end{array}
\end{equation}

\section{Classical Bayesian nets and disintegration}\label{sec:disintegration}

A major rationale for using Bayesian
networks~\cite{Pearl88,BernardoS00,Barber12,KollerF09} is efficiency
of representation: a joint probability distribution (state) on
multiple sample spaces (domains) quickly becomes very
large. Representing the same distribution in graphical form, as a
`Bayesian network' is often much more efficient. The directed graph
structure is determined by conditional independence. Semantically, the
directed arcs are given by channels, that is by stochastic matrices,
or more abstractly by Kleisli morphisms for the distribution
monad~\cite{Fong12,JacobsZ16,JacobsZ17}.

The essence of this semantical view on Bayesian network theory
consists of two parts.
\begin{enumerate}[label=(\Roman*)]
\item \label{essencebayesjoint} The ability to move back-and-forth
  between a joint state and a graph (network) of channels. The
  difficult direction is extracting the various channels of the graph
  from a joint state. This is often called
  \emph{disintegration}~\cite{ChoJ17a}.

\item \label{essencebayesinference} Equivalence of inference via joint
  states and inference via associated channels. In general, inference
  (or, Bayesian learning) happens via conditioning (updating,
  revising) of states, in the light of evidence given by
  predicates. Inference involves the propagation of such conditioning
  via joint states and/or via channels, via the bank-and-forth
  connections in~\ref{essencebayesjoint}, both in a forward and
  backward direction (as in~\cite{JacobsZ16,JacobsZ18}).
\end{enumerate}

\noindent Point~\ref{essencebayesjoint} is well-known, but
point~\ref{essencebayesinference} is usually left implicit; it is
however a crucial part of why efficient representation of (big) joint
states as Bayesian network graphs can be used for Bayesian
reasoning. In this section we briefly elaborate both points below, and
illustrate them with an example.

Note that we do not claim that with these two
points~\ref{essencebayesjoint} and~\ref{essencebayesinference} we
capture all essentials of Bayesian network theory: \textit{e.g.}, we
do not address the matter of \emph{how} to turn a joint state into a
graph, via conditional independence or via causality. This question
has also be studied in a quantum setting, see
\textit{e.g.}~\cite{PienaarB15}.

\subsection{Disintegration}\label{subsec:disintegration}

\begin{wrapfigure}[5]{r}{0pt}
\begin{minipage}{11em}\centering
\vspace*{-1.5em}
\begin{equation}
\label{diag:pairing}
\hspace*{-1em}\pair(\omega,c) 
\coloneqq 
\vcenter{\hbox{%
\begin{tikzpicture}[font=\small,scale=0.7]
\node[state] (s) at (0,0) {$\omega$};
\node[copier] (copier) at (0,0.3) {};
\node[arrow box] (c) at (0.5,0.95) {$c$};
\coordinate (X) at (-0.5,1.5);
\coordinate (Y) at (0.5,1.5);
\draw (s) to (copier);
\draw (copier) to[out=150,in=-90] (X);
\draw (copier) to[out=15,in=-90] (c);
\draw (c) to (Y);
\end{tikzpicture}}}
\end{equation}
\end{minipage}
\end{wrapfigure}
Abstractly, point~\ref{essencebayesjoint} involves the correspondence
between a joint state on the one hand, and a channel and a (single)
state on the other hand.  In one direction this is easy: given a state
$\omega$ on $X$ and a channel $c\colon X \rightarrow Y$ we can form a
joint state on $X\times Y$, namely as: $\pair(\omega, c) \coloneqq
\big((\idmap\otimes c) \klafter \Delta\big) \gg \omega$, where $\Delta
\colon X \rightarrow X\times X$ is the copier channel with $\Delta(x)
= 1\ket{x,x}$. This construction is drawn as a picture on the
right~\eqref{diag:pairing}, using the graphical language associated
with monoidal categories. It will be used here only as illustration,
hopefully in an intuitive self-explanatory manner. We refer
to~\cite{Selinger11,CoeckeK16,ChoJ17a} for details.

\begin{wrapfigure}[6]{r}{0pt}
\begin{minipage}{13em}\centering
\begin{equation}
\label{diag:disintegration}
\vcenter{\hbox{%
\begin{tikzpicture}[font=\small,scale=0.7]
\node[state] (omega) at (0,0) {$\,\;\tau\;\,$};
\coordinate (X) at (-0.4,0.55) {};
\coordinate (Y) at (0.4,0.55) {};
\draw (omega) ++(-0.4, 0) to (X);
\draw (omega) ++(0.4, 0) to (Y);
\end{tikzpicture}}}
\;
=
\;
\vcenter{\hbox{%
\begin{tikzpicture}[font=\small]
\node[state] (omega) at (0.25,0) {$\,\;\tau\;\,$};
\node[copier] (copier) at (0,0.4) {};
\node[arrow box] (c) at (0.5,0.95) {$c$};
\coordinate (X) at (-0.5,1.5);
\coordinate (Y) at (0.5,1.5);
\coordinate (omega1) at ([xshiftu=-0.25]omega);
\coordinate (omega2) at ([xshiftu=0.25]omega);
\node[discarder] (d) at ([yshiftu=0.2]omega2) {};
\draw (omega1) to (copier);
\draw (omega2) to (d);
\draw (copier) to[out=150,in=-90] (X);
\draw (copier) to[out=15,in=-90] (c);
\draw (c) to (Y);
\end{tikzpicture}}}
\end{equation}
\end{minipage}
\end{wrapfigure}
Going in the other direction, from a joint state to a channel is less
trivial. It is called \emph{disintegration} \textit{e.g.}
in~\cite{ChoJ17a}. It involves a joint state $\tau$ on $X,Y$ from
which a channel $c\colon X \rightarrow Y$ is extracted, in such a way
that $\tau$ itself can be reconstructed from its first marginal
$\margsign_{1}(\tau)$ and the channel $c$. Pictorially this marginal
is represented by blocking its second wire via the ground symbol
$\ground$. We write $\extract(\tau)$ for this extracted channel
$c$. Then we can write Equation~\eqref{diag:disintegration} as $\tau =
\pair\big(\margsign_{1}(\tau), \extract(\tau)\big)$.


\begin{lemma}
\label{lem:disintegration}
Extracted channels $\extract(\tau)$ exist and are unique in classical
discrete probability, for joint states $\tau$ whose first marginal has
full support.
\end{lemma}

\begin{myproof}
First, a state $\omega\in\Dst(X)$ and a channel $c\colon X \rightarrow
\Dst(Y)$ yield a joint state $\pair(\omega, c) \in \Dst(X\times Y)$,
namely, as described in~\eqref{diag:pairing} above:
\[ \begin{array}{rcccl}
\pair(\omega, c)(x,y)
& = &
\big(((\idmap\otimes c) \klafter \Delta) \gg \omega\big)(x,y)
& = &
\omega(x) \cdot c(x)(y).
\end{array} \]

\noindent In the other direction, let $\tau\in\Dst(X\times Y)$ be a
joint state whose first marginal $\margsign_{1}(\tau) \in \Dst(X)$ has
full support. The latter means that its support is the whole of $X$,
so that: $\margsign_{1}(\tau)(x) \neq 0$ for each $x$, or,
equivalently, $\all{x}{\ex{y}{\tau(x,y)\neq 0}}$. We can now define a
channel $\extract(\tau) \colon X \rightarrow \Dst(Y)$ by:
\[ \begin{array}{rcccl}
\extract(\tau)(x)(y)
& = &
\displaystyle\frac{\tau(x,y)}{\margsign_{1}(\tau)(x)}
& = &
\displaystyle\frac{\tau(x,y)}{\sum_{z} \tau(x,z)}.
\end{array} \eqno{\QEDbox} \]
\end{myproof}

For a more systematic, diagrammatic description of disintegration,
also for continuous probability, we refer to~\cite{ChoJ17a}. Here we
only need it for discrete probability, as a preparation for the
quantum case.

\subsection{Excursion on disintegration and 
   semi-exponentials}\label{subsec:excursion}

We conclude this part on disintegration with a novel observation. It
is interesting in itself, but it does not play a role in the
sequel. It shows that disintegration gives rise to higher order
`semi-exponential' structure, originally introduced
in~\cite{Hayashi85}. Recall that a categorical description of a
(proper) exponential in a cartesian closed category involves exponent
objects $Y^X$ with an evaluation map $\evmap \colon Y^{X} \times X
\rightarrow Y$ such that for each map $f\colon Z\times X \rightarrow
Y$ there is an abstraction map $\Lambda(f) \colon Z \rightarrow
Y^{X}$. These $\evmap$ and $\Lambda$ should satisfy:
\begin{equation}
\label{eqn:exponential}
\begin{array}{rclcrclcrcl}
\evmap \after (\Lambda(f)\times\idmap)
& = &
f
& \qquad &
\Lambda(f \after (g\times\idmap))
& = &
\Lambda(f) \after g
& \qquad &
\Lambda(\evmap)
& = &
\idmap.
\end{array}
\end{equation}

\noindent The last two equations ensure that $\Lambda(f)$ is the
unique map $h$ with $\evmap \after (h\times\idmap) = f$, since: $h =
\idmap \after h = \Lambda(\evmap) \after h = \Lambda(\evmap \after
(h\times\idmap)) = \Lambda(f)$.

For a \emph{semi-exponential}, the last equation
in~\eqref{eqn:exponential} need not hold. A semi-exponential is thus
more than a `weak' exponential (only the first equation) since it also
satisfies naturality (the second equation).  In the language of the
$\lambda$-calculus, having `semi-exponentials' means that one has a
$\beta$-equation, but not an $\eta$-equation, see~\cite{Hayashi85}
or~\cite{HoofmanM95} for more details.

\begin{theorem}
\label{thm:semiexponential}
Let $\Kl_{\mathsf{f}}(\Dst)$ be the subcategory of the Kleisli
category $\Kl(\Dst)$ of the distribution monad on $\Sets$ with only
\emph{finite} sets as objects. This category $\Kl_{\mathsf{f}}(\Dst)$
is symmetric monoidal `semi' closed: it has semi-exponentials, which
are semi-right adjoint to the (standard) tensor product.
\end{theorem}

\begin{myproof}
It is well-known that cartesian products $\times$ on sets and parallel
product $\otimes$ on Kleisli maps (channels) makes the category
$\Kl(\Dst)$, and also $\Kl_{\mathsf{f}}(\Dst)$, symmetric monoidal
closed.  We sketch how semi-exponentials $\multimap$ are obtained via
joint states whose first marginal has full support:
\[ \begin{array}{rcl}
X \multimap Y
& \coloneqq &
\setin{\tau}{\Dst(X\times Y)}{\supp\big(\margsign_{1}(\tau)\big) = X}.
\end{array} \]

\noindent This definition assumes that $X$ is not the empty set. In
that case we can set $\emptyset\multimap Y = 1$, the singleton set,
since $Z\times \emptyset \cong \emptyset$ so that there is a trivial
correspondence between maps $Z\times\emptyset\rightarrow\Dst(Y)$ and
maps $Z \rightarrow \Dst(1) = 1$.

We define an evaluation channel $\evmap \colon (X\multimap Y)\times X
\rightarrow \Dst(Y)$ via disintegration:
\[ \begin{array}{rcccl}
\evmap(\tau, x)(y)
& \coloneqq &
\extract(\tau)(x)(y)
& = &
\displaystyle\frac{\tau(x,y)}{\sum_{z}\tau(x,z)}.
\end{array} \]

\noindent For abstraction, let $f\colon Z\times X \rightarrow \Dst(Y)$
be given. We define $\Lambda(f) \colon Z \rightarrow \Dst(X\multimap
Y)$ as:
\[ \begin{array}{rclcrcl}
\Lambda(f)(z)
& \coloneqq &
1\bigket{\tau}
&\qquad \mbox{where} \qquad &
\tau(x,y)
& = &
\displaystyle\frac{f(z,x)(y)}{\# X},
\end{array} \]

\noindent and where $\# X$ is the number of elements in $X$. Here we
construct a joint state $\tau = \pair(\uniform, f(z, -))$ as in the
beginning of this section, from the uniform distribution $\uniform$ on
$X$ and the channel $f(z,-)\colon X \rightarrow \Dst(Y)$. We need to
check that $\Lambda(f)$ is well-defined, in particular that each first
marginal of $\Lambda(f)(z)(*) \in \Dst(X\times Y)$ has full support:
\[ \begin{array}{rcccccccl}
\margsign_{1}(\Lambda(f)(z)(*))(x)
& = &
\sum_{y} \Lambda(f)(z)(*)(x,y)
& = &
\sum_{y} \frac{f(z,x)(y)}{\# X}
& = &
\frac{1}{\# X}
& \neq &
0.
\end{array} \]

\noindent It is easy to check that the first two equations
from~\eqref{eqn:exponential} hold. \QED

\auxproof{
We first check that the construction of $\tau$ is indeed a pair,
see the first equation in the proof of Lemma~\ref{lem:disintegration}:
\[ \begin{array}{rcccccl}
\pair(\uniform, f(z,-))(x,y)
& = &
\uniform(x) \cdot f(z,x)(y)
& = &
\displaystyle\frac{f(z,x)(y)}{\# X}
& = &
\tau(x,y).
\end{array} \]

\noindent Next we look at the required equations:
\[ \begin{array}{rcl}
\big(\evmap \klafter (\Lambda(f) \otimes\idmap)\big)(z,x)(y)
& = &
\evmap(\Lambda(f)(z)(*), x)(y)
\\
& = &
\displaystyle\frac{\Lambda(f)(z)(x,y)}{\sum_{y'}\Lambda(f)(z)(x,y')}
\\
& = &
\displaystyle\frac{\nicefrac{f(z,x)(y)}{\# X}}
                  {\sum_{y'}\nicefrac{f(z,x)(y')}{\# X}}
\\
& = &
f(z,x)(y).
\end{array} \]

\noindent Next, for $g\colon W \rightarrow \Dst(Z)$,
\[ \begin{array}{rcl}
\Lambda(f \klafter (g\otimes\idmap))(w)(*)(x,y)
& = &
\displaystyle\frac{(f \klafter (g\otimes\idmap))(w,x)(y)}{\# X}
\\
& = &
\displaystyle\frac{\sum_{z} g(w)(z) \cdot f(z,x)(y)}{\# X}
\\
& = &
\sum_{z} g(w)(z) \cdot \displaystyle\frac{f(z,x)(y)}{\# X}
\\
& = &
\sum_{z} g(w)(z) \cdot \Lambda(f)(z)(*)(x,y)
\\
& = &
\big(\Lambda(f) \klafter g\big)(w)(*)(x,y).
\end{array} \] 
}
\end{myproof}

\subsection{Bayesian inference and disintegration}\label{subsec:inference}

We now turn to the second point~\ref{essencebayesinference} from the
very beginning of this section, about Bayesian inference, especially
in relation to the passage back-and-forth between joint states and
channels via pairing and extraction, as just described.

It may happen that a joint state $\tau\in\Dst(X\times Y)$ is equal to
the product of its two marginals, \textit{i.e.}~$\tau =
\margsign_{1}(\tau) \otimes \margsign_{2}(\tau)$. The state $\tau$ is
then called \emph{non-entwined}. The more common case is that a joint
state is entwinted, and its marginal components are correlated.  If we
then update in one component, we see a change in the other
component. This is called \emph{crossover influence}
in~\cite{JacobsZ17,JacobsZ18}.

The essence of the point~\ref{essencebayesinference}, in the beginning
of this section, about inference and disintegration is that for a
joint state $\tau$, this crossover influence can be propagated through
the channel $c$ that is extracted from the state $\tau$ via
disintegration. This is expressed in the next result, called the
\emph{Bayesian Inference Theorem}.

\begin{theorem}
\label{thm:classicalconditioning}
Let $\tau\in\Dst(X\times Y)$ be a joint state, and $c = \extract(\tau)
\colon X \rightarrow \Dst(Y)$ the extracted channel obtained via
disintegration --- as described in
Subsection~\ref{subsec:disintegration}. For predicates $p\in
[0,1]^{X}$ and $q\in [0,1]^{Y}$ we then have:
\begin{equation}
\label{eqn:classicalconditioning}
\begin{array}{rclcrcl}
\margsign_{2}\big(\tau|_{p\otimes\one}\big)
& = &
c \gg \big(\margsign_{1}(\tau)\big|_{p}\big)
& \qquad\mbox{and}\qquad &
\margsign_{1}\big(\tau|_{\one\otimes q}\big)
& = &
\margsign_{1}(\tau)\big|_{c \ll q}.
\end{array}
\end{equation}
\end{theorem}

The first equation describes crossover inference on the left-hand-side
as \emph{forward} inference on the right: first update and then do
state transformation $\gg$. The second equation
in~\eqref{eqn:classicalconditioning} describes crossover inference in
the other component as \emph{backward} inference: first do predicate
transformation $\ll$ and then update. The terminology of `forward' and
`backward' inference comes from~\cite{JacobsZ16}, see
also~\cite{JacobsZ18}.  An abstract graphical proof of the
equations~\eqref{eqn:classicalconditioning} is given
in~\cite{ChoJ17a}. But it is not hard to prove these equations
concretely, by unwrapping the definitions.

\auxproof{
\begin{proof}
Essentially this is a matter of unwrapping the relevant definitions.
We do the first case:
\[ \begin{array}[b]{rcl}
\big(c \gg ((\marg{\tau}{[1,0]})|_{p})\big)(y)
& = &
{\displaystyle\sum}_{x} c(x)(y) \cdot ((\marg{\tau}{[1,0]})|_{p})(x) 
\\[+0.2em]
& = &
{\displaystyle\sum}_{x} c(x)(y) \cdot \displaystyle
   \frac{(\marg{\tau}{[1,0]})(x)\cdot p(x)}{\marg{\tau}{[1,0]}\models p}
\\[+0.8em]
& \smash{\stackrel{\eqref{diag:disintegration}}{=}} &
{\displaystyle\sum}_{x} \displaystyle
   \frac{\tau(x,y)\cdot p(x)}{\sum_{x'}(\marg{\tau}{[1,0]})(x')\cdot p(x')}
\\[+0.8em]
& = &
{\displaystyle\sum}_{x} \displaystyle
   \frac{\tau(x,y)\cdot (p\otimes\one)(x,y)}{\sum_{x',y'}(\tau(x',y')\cdot p(x')}
\\[+0.8em]
& = &
{\displaystyle\sum}_{x} \displaystyle
   \frac{\tau(x,y)\cdot (p\otimes\one)(x,y)}{\tau \models p\otimes\one}
\\[+0.6em]
& = &
{\displaystyle\sum}_{x} \tau|_{p\otimes\one}(x,y)
\\[+0.2em]
& = &
\big(\marg{(\tau|_{p\otimes\one})}{[0,1]}\big)(y).
\end{array} \eqno{\QEDbox} \]

\auxproof{
\[ \begin{array}{rcl}
\big((\marg{\tau}{[1,0]})|_{c \ll q}\big)(x)
& = &
\displaystyle\frac{(\marg{\tau}{[1,0]})(x) \cdot (c \ll q)(x)}
                  {\marg{\tau}{[1,0]} \models c \ll q}
\\[+0.8em]
& = &
{\displaystyle\sum}_{y} \displaystyle
   \frac{(\marg{\tau}{[1,0]})(x) \cdot c(x)(y) \cdot q(y)}
        {\sum_{x'} (\marg{\tau}{[1,0]})(x') \cdot (c \ll q)(x')}
\\[+0.8em]
& \smash{\stackrel{\eqref{diag:disintegration}}{=}} &
{\displaystyle\sum}_{y} \displaystyle
   \frac{\tau(x,y) \cdot q(y)}
        {\sum_{x',y'} (\marg{\tau}{[1,0]})(x') \cdot c(x')(y') \cdot q(y')}
\\[+0.8em]
& \smash{\stackrel{\eqref{diag:disintegration}}{=}} &
{\displaystyle\sum}_{y} \displaystyle
   \frac{\tau(x,y) \cdot q(y)}{\sum_{x',y'}\tau(x',y') \cdot q(y')}
\\[+0.8em]
& = &
{\displaystyle\sum}_{y} \displaystyle
   \frac{\tau(x,y) \cdot (\one\otimes q)(x,y)}{\tau \models \one\otimes q}
\\[+0.8em]
& = &
{\displaystyle\sum}_{y} \tau|_{(\one\otimes q)}(x,y)
\\[+0.2em]
& = &
\big(\marg{\tau|_{(\one\otimes q)}}{[1,0]}(x).
\end{array} \]
}
\end{proof}
}

\subsection{An illustration of inference in a classical 
   Bayesian network}\label{subsec:smoking}

We consider the relation between smoking and the presence of ashtrays
and (lung) cancer, in the following simple Bayesian network.
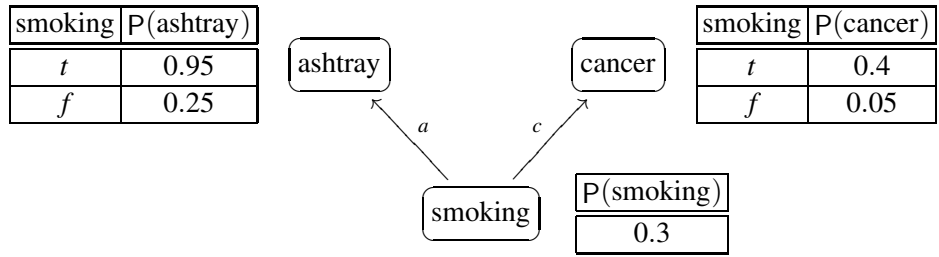
\[ \xymatrix@C-1.5pc@R-1.5pc{
{\setlength\tabcolsep{0.2em}\renewcommand{\arraystretch}{1}
\begin{tabular}{|c|c|}
\hline
smoking & $\Prob(\text{ashtray})$ \\
\hline\hline
$t$ & $0.95$ \\
\hline
$f$ & $0.25$ \\
\hline
\end{tabular}} 
& \ovalbox{\strut ashtray} & & \ovalbox{\strut cancer} &
{\setlength\tabcolsep{0.2em}\renewcommand{\arraystretch}{1}
\begin{tabular}{|c|c|}
\hline
smoking & $\Prob(\text{cancer})$ \\
\hline\hline
$t$ & $0.4$ \\
\hline
$f$ & $0.05$ \\
\hline
\end{tabular}}
\\
\\
& & \ovalbox{\strut smoking}\ar[uul]_{a}\ar[uur]^{c}
  \rlap{\quad\smash{\setlength\tabcolsep{0.2em}\renewcommand{\arraystretch}{1}
\begin{tabular}{|c|}
\hline
$\Prob(\text{smoking})$ \\
\hline\hline
$0.3$ \\
\hline
\end{tabular}}} 
}\]

\noindent Thus, 95\% of people who smoke have an ashtray in their
home, and 25\% of the non-smokers too. On the right we see that in
this situation a smoker has 40\% chance of developing cancer, whereas
a non-smoker only has 5\% chance.

The question we want to address is: what is the influence of the
presence or absence of an ashtray on the probability of developing
cancer? Here the presence/absence of the ashtray is the `evidence',
whose influence is propagated through the network. We shall describe
the outcome using the EfProb tool~\cite{ChoJ17b}, concentrating on
evidence propagation, and not so much on the precise representation of
the above network, using channels \pythoninline{a} and
\pythoninline{c} associated with the conditional probability tables.

We first consider the prior probabilities of smoking,
ashtray, and cancer:
\begin{python}
>>> smoking
0.3|t> + 0.7|f>
>>> a >> smoking
0.46|t> + 0.54|f>
>>> c >> smoking
0.155|t> + 0.845|f>
\end{python}

\noindent The network gives rise to a joint state, by tupling the
ashtray, identity and cancer channels, and applying them to the
\pythoninline{smoking} state.  We can then obtain the above three
prior probabilities alternatively via three marginalisations of this
joint state, namely as first, second, third marginals, by using in
EfProb the corresponding masks \pythoninline{[1,0,0]},
\pythoninline{[0,1,0]}, \pythoninline{[0,0,1]} after the
marginalisation sign \pythoninline{\%}.
\begin{python}
>>> joint = (a @ idn(bnd) @ c) * copy(bnd,3) >> smoking
>>> joint
0.114|t,t,t> + 0.171|t,t,f> + 0.00875|t,f,t> + 0.166|t,f,f> 
 + 0.006|f,t,t> + 0.009|f,t,f> + 0.0263|f,f,t> + 0.499|f,f,f>
>>> joint % [1,0,0]
0.46|t> + 0.54|f>
>>> joint % [0,1,0]
0.3|t> + 0.7|f>
>>> joint % [0,0,1]
0.155|t> + 0.845|f>
\end{python}

\noindent We now wish to infer the (adapted) cancer probability when
we have evidence of ashtrays. We shall do this in two ways, first via
crossover inference using the above joint state. The ashtray evidence
\pythoninline{tt} needs to be extended (weakened) to a predicate with
the same domain as the joint state. In the
Equations~\eqref{eqn:classicalconditioning} this is written as:
$p\otimes\one$, but in EfProb it is: \pythoninline{tt @ truth(bnd) @
  truth(bnd)}. We first use this predicate for updating the joint
state, written as \pythoninline{/} in EfProb, and then we marginalise
to obtain the third `cancer' component that we are interested in:
\begin{python}
>>> (joint / (tt @ truth(bnd) @ truth(bnd))) % [0,0,1]
0.267|t> + 0.733|f>
\end{python}

\noindent Alternatively we can compute this posterior cancer
probability by following the graph structure. The ashtray evidence
\pythoninline{tt} is now first turned into predicate \pythoninline{a
  << tt} on the state \pythoninline{smoking}. After updating this
state, we transform it to an updated cancer probability, via state
transformation \pythoninline{>>}. We can do this down-and-up
propagation in one go:
\begin{python}
>>> c >> (smoking / (a << tt))
0.267|t> + 0.733|f>
\end{python}

\noindent The fact that we get the same distribution is an instance of
the equations~\eqref{eqn:classicalconditioning}. As expected, in
presence of ashtrays the probability of cancer is higher.

Aside: clearly, ashtrays \emph{influence} (the probability of) cancer,
but they are not the \emph{cause}; in the graph this influence
happens via a common ancestor, namely smoking, working 
statistically as `confounder', and as the actual cause of cancer.

\section{Towards quantum Bayesian theory}\label{sec:quantum}

The main aim of this paper is to investigate quantum analogues of the
Bayesian Inference Theorem~\ref{thm:classicalconditioning}, from the
conviction that any adequate quantum Bayesian network theory should
address these points~\ref{essencebayesjoint}
and~\ref{essencebayesinference} from the beginning of
Section~\ref{sec:disintegration} in a satisfactory
manner. Point~\ref{essencebayesjoint} has received ample attention in
quantum theory, see for
instance~\cite{LeiferS13,CoeckeS12,LeiferS14,AllenBHLS17}. But
Point~\ref{essencebayesinference} involving quantum conditioning has
not really been studied this explicitly. Our main result is that one
can also describe quantum conditioning consistently, both on joint
states and via channels, as in
Equations~\eqref{eqn:classicalconditioning}, but this requires in the
quantum case that one distinguishes \emph{two forms of conditioning},
which we shall call \emph{lower} and \emph{upper} conditioning,
written as $\sigma|_{p}$ and $\sigma|^{p}$ respectively\footnote{The
  terminology `lower' and `upper' is simply determined by the position
  of the predicate $p$, low in $\sigma|_{p}$ and up in
  $\sigma|^{p}$.}. Classically these two forms of conditioning
coincide, but the quantum world is more subtle --- as usual. Lower
conditioning has appeared in effectus theory~\cite{ChoJWW15b} and
upper conditioning in the approach of~\cite{LeiferS13}.  Here they are
clearly distinguished for the first time, and used jointly to capture
quantum inference and propagation of evidence.  Interestingly, what is
commonly called Bayes' rule holds for upper conditioning, but not for
lower conditioning, for which we ``only'' have the product rule.

First we introduce the basics about states and predicates in the
quantum world. We shall do so for finite-dimensional quantum theory,
using the formalism of Hilbert spaces.

\subsection{Basics of quantum probability}\label{subsec:quantum}


Let $\mathscr{H}$ be a finite-dimensional complex Hilbert space. A
\emph{state} $\sigma$ of $\mathscr{H}$ is a positive operator on
$\mathscr{H}$ with trace one. That is, $\sigma$ is linear function
$\sigma \colon \mathscr{H} \rightarrow \mathscr{H}$ satisfying $\sigma
\geq 0$ and $\tr(\sigma) = 1$.  A state is often called a
\emph{density matrix}. The canonical way to define a state is to start
from a vector $\ket{v} \in \mathscr{H}$ with norm $1$, and consider
the operator $\ket{v}\bra{v} \colon \mathscr{H} \rightarrow
\mathscr{H}$. It sends any element $\ket{w}\in\mathscr{H}$ to the
vector $\tuple{v|w}\cdot\ket{v}$. An arbitrary state is a convex
combination of such vector states $\ket{v}\bra{v}$. A \emph{joint}
state $\tau$ on two Hilbert space $\mathscr{H}$ and $\mathscr{K}$ is a
state on the tensor product $\mathscr{H}\otimes\mathscr{K}$.


A \emph{predicate}, also called an \emph{effect}, is a positive
operator $p$ on $\mathscr{H}$ below the identity: $0 \leq p \leq
\idmap$. The identity $\idmap$ is given by the identity/unit matrix,
and corresponds to the \emph{truth} predicate, often written as
$\one$. For each predicate $p$ there is an orthosupplement, written as
$p^\bot$, playing the role of negation. It is defined by $p^{\bot} =
\idmap - p$, and satisfies: $p^{\bot\bot} = p$ and $p + p^{\bot} =
\one$.

The most interesting logical operation on quantum predicates is
\emph{sequential conjunction} $\andthen$. It is defined via the square
root operation on predicates, as:
\begin{equation}
\label{AndthenEqn}
\begin{array}{rcl}
p \andthen q
& = &
\sqrt{p}\,q\sqrt{p}.
\end{array}
\end{equation}
\noindent We pronounce $\andthen$ as `and-then', and read it as: after
$p$ with its side-effect, the predicate $q$ holds. This operation
$\andthen$ has been studied in~\cite{GudderG02}, and re-emerged in
effectus theory~\cite{Jacobs15d,ChoJWW15b}. The square root of the
matrix $p$ exists since $p$ is positive. It is computed via
diagonalisation $\sqrt{p} = U\sqrt{D}U^{-1}$, where $p = UDU^{-1}$, in
which $\sqrt{D}$ is obtained from the diagonal matrix $D$ by taking
the square roots of the (positive) eigenvalues on the diagonal.

States $\sigma$ and predicates $p$ of the same Hilbert space
$\mathscr{H}$ can be combined in \emph{validity}, defined as:
\begin{equation}
\label{ValidityEqn}
\begin{array}{rcl}
\sigma\models p
& \;\coloneqq\; &
\tr(\sigma\,p) \;\in\; [0,1].
\end{array}
\end{equation}
\noindent This standard definition is also known as the Born rule.


\begin{rem}
\label{rem:classicaltoquantum}
There is a standard way to embed classical probability into quantum
probability. Suppose we have classical state $\omega$ and predicate
$p$ on a finite set $X = \{x_{1}, \ldots, x_{n}\}$ with $n$
elements. Then we consider the Hilbert space $\C^{n}$ with standard
basis given by vectors $\ket{i}$ with an $1$ on the $i$-th position
and zeros elsewhere. We write $\widehat{\omega} = \sum_{i}
\omega(x_{i})\ket{i}\bra{i}$ for the `diagonal' quantum state on
$\C^n$. By construction it is positive and has trace
$\sum_{i}\omega(x_{i}) = 1$.

Similarly, a classical predicate $p \in [0,1]^{X}$ gives a quantum
predicate $\widehat{p}$ on $\C^{n}$ via $\widehat{p} = \sum_{i}
p(x_{i})\ket{i}\bra{i}$. By construction, $0 \leq p \leq \idmap$.  It
is easy to see that the classical and quantum validities coincide:
\[ \begin{array}{rcccccl}
\omega\models p
& = &
\sum_{i} \omega(i)\cdot p(i)
& = &
\tr\big(\widehat{\omega}\,\widehat{p}\big)
& = &
\widehat{\omega} \models \widehat{p}.
\end{array} \]

\noindent The mapping $\smash{\widehat{(\;\cdot\;)}}$ preserves the
logical structure on predicates, including sequential conjunction
$\andthen$.
\end{rem}

\begin{rem}
\label{rem:stateispredicate}
In both classical and quantum probability, as described here, a state
is also a predicate. This is peculiar. When one moves to a higher level
of abstraction, this is no longer the case --- for instance by using
von Neumann algebras instead of Hilbert spaces, or by using continuous
probability distributions on measurable spaces instead of discrete
distributions on sets. In the next section we sometimes `convert' a
state into a predicate, but we shall make explicit when we do so.  A
more abstract approach is possible, using the duality between states
and effects, see also Remark~\ref{rem:daggerconditioning}.
\end{rem}

\subsection{Two forms of quantum conditioning}\label{subsec:quantumconditioning}

This subsection introduces two forms of quantum conditioning of a
state by a predicate, called `lower' and `upper' conditioning, and
describes their basic properties.

\begin{definition}
\label{def:quantumconditioning}
Let $\sigma$ be a state, and $p$ a predicate, on the same Hilbert
space, for which the validity $\sigma\models p$ is non-zero.
We shall use the following terminology, notation and definition
for two forms of conditioning:
\[ \begin{array}{rclcrcl} 
\mbox{lower:}\quad 
\sigma|_{p}
& \coloneqq &
\displaystyle\frac{\sqrt{p}\,\sigma\sqrt{p}}{\sigma\models p}
& \qquad\qquad &
\mbox{upper:}\quad 
\sigma|^{p}
& \coloneqq &
\displaystyle\frac{\sqrt{\sigma}\,p\sqrt{\sigma}}{\sigma\models p}.
\end{array} \]
\end{definition}

It is easy to see that both $\sigma|_{p}$ and $\sigma|^{p}$ are states
again --- using the familiar `rotation' property of traces: $\tr(AB) =
\tr(BA)$. Lower conditioning $\sigma|_{p}$ arises in effectus theory,
whereas upper conditioning $\sigma|^{p}$ comes
from~\cite{LeiferS13}. We first observe that this difference between
`lower' and `upper' does not exist classically.

\begin{lemma}
\label{lem:quantumconditioningclassical}
For classical (non-quantum) states and predicates, lower and upper
conditioning coincide with classical conditioning. To express this
more precisely we use the notation $\smash{\widehat{(\;\cdot\;)}}$
from Remark~\ref{rem:classicaltoquantum} to translate from classical
to quantum: for a classical state $\omega$ and predicate $p$,
\[ \begin{array}{rcccl}
\widehat{\omega}|_{\widehat{p}}
& = &
\omega|_{p}
& = &
\widehat{\omega}|^{\widehat{p}}.
\end{array} \]
\end{lemma}

\begin{myproof}
Diagonal matrices commute, so that
$\sqrt{\widehat{p}}\;\widehat{\omega}\,\sqrt{\widehat{p}} =
\widehat{\omega}\,\widehat{p} =
\sqrt{\widehat{\omega}}\,\widehat{p}\,\sqrt{\widehat{\omega}}$. \QED
\end{myproof}

A second observation is about truth $\one$ and sequential conjunction
$\andthen$. Both lower and upper conditioning with truth $\one$ does
nothing, like in the classical case, but successive conditioning
cannot be reduced to single conditioning, like in the first equation
in~\eqref{eqn:classicalsuccesiveconditioning}, in
Proposition~\ref{prop:classicalbayes}. In addition, the order in
quantum conditioning matters, just like the order of priming in
psychology matters~\cite{Jacobs17f}.

\begin{rem}
\label{rem:quantumconditioningconjunctions}
We have $\sigma|_{\one} = \sigma$ and $\sigma|^{\one} = \sigma$, but
in general successive quantum conditionings cannot be reduced to
a single conditioning via sequential conjunction:
\[ \begin{array}{rclcrcl}
(\sigma|_{p})|_{q} 
& \neq &
\sigma|_{p\andthen q}
& \qquad\mbox{and also}\qquad &
(\sigma|^{p})|^{q} 
& \neq &
\sigma|^{p\andthen q}.
\end{array} \]

\noindent Similarly, in general, quantum conditionings do not commute:
\[ \begin{array}{rclcrcl}
(\sigma|_{p})|_{q} 
& \neq &
(\sigma|_{q})|_{p}
& \qquad\mbox{and}\qquad &
(\sigma|^{p})|^{q} 
& \neq &
(\sigma|^{q})|^{p}.
\end{array} \]
\end{rem}

Interestingly, the two classical equations~\eqref{eqn:classicalbayes}
in Proposition~\ref{prop:classicalbayes} hold separately for the two
kinds of quantum conditioning.

\begin{proposition}
\label{prop:quantumbayes}
The `product' rule holds for lower conditioning and Bayes' rule holds
for upper conditioning:
\begin{equation}
\label{eqn:quantumbayes}
\begin{array}{rclcrcl}
\sigma|_{p} \models q
& = &
\displaystyle\frac{\sigma\models p\andthen q}{\sigma\models p}
& \hspace*{5em} &
\sigma|^{p} \models q
& = &
\displaystyle\frac{(\sigma|^{q}\models p)\cdot(\sigma\models q)}{\sigma\models p}.
\end{array}
\end{equation}
\end{proposition}

\begin{myproof}
We simply go through the computations:
\[ \mystretch\begin{array}[b]{rcccccl}
\sigma|_{p} \models q
& = & 
\tr\big(\sigma|_{p}\,q\big)
& = &
\tr\big(\frac{\sqrt{p}\,\sigma\sqrt{p}}{\sigma\models p}\,q\big)
& = &
\frac{1}{\sigma\models p}\cdot\tr\big(\sigma\sqrt{p}\,q\sqrt{p}\big)
\hspace*{\arraycolsep}=\hspace*{\arraycolsep}
\frac{\sigma\models p\andthen q}{\sigma\models p}
\\[+0.3em]
\sigma|^{p} \models q
& = & 
\tr\big(\sigma|^{p} \, q\big)
& = &
\tr\big(\frac{\sqrt{\sigma}\,p\sqrt{\sigma}}{\sigma\models p}\,q\big)
& = &
\frac{\sigma\models q}{\sigma\models p}\cdot 
   \tr\big(p\frac{\sqrt{\sigma}\,q\sqrt{\sigma}}{\sigma\models q}\big)
\hspace*{\arraycolsep}=\hspace*{\arraycolsep}
\frac{(\sigma|^{q}\models p)\cdot(\sigma\models q)}{\sigma\models p}.
\end{array} \eqno{\QEDbox} \]
\end{myproof}

\section{Quantum channels}\label{sec:quantumchannel}

In order to express the quantum analogues of the equations in
Theorem~\ref{thm:classicalconditioning} we need the notion of
`channel' in a quantum setting. It exists, and is alternatively often
called a quantum operation, see \eg~\cite{NielsenC00}. There are
several variations possible in the requirements, such as just positive
or complete positive, unitary or subunitary, normal or not. These
variations are not essential for what follows.

For a finite-dimensional Hilbert space $\mathscr{H}$ be write
$\mathcal{B}(\mathscr{H})$ for the set of linear maps $A\colon
\mathscr{H} \rightarrow \mathscr{H}$. Because $\mathscr{H}$ has finite
dimension, such $A$ are automatically bounded, or equivalently,
continuous. The set of operators $\mathcal{B}(\mathscr{H})$ is in fact
a Hilbert space itself, with Hilbert-Schmidt inner product
$\HSip{A}{B} = \tr(A^{\dag}B)$, where $A^{\dag}$ is the conjugate
transpose of $A$, as matrix. Moreover, there are canonical
isomorphisms $\mathcal{B}(\mathscr{H}\otimes\mathscr{K}) \cong
\mathcal{B}(\mathscr{H}) \otimes \mathcal{B}(\mathscr{K})$ and
$\mathcal{B}(\C) \cong \C$.

If $\mathscr{K}$ is another finite-dimensional Hilbert space, then a
\emph{CP-map} $\mathscr{H} \rightarrow \mathscr{K}$ is a completely
positive linear map $c\colon \mathcal{B}(\mathscr{K}) \rightarrow
\mathcal{B}(\mathscr{H})$. Notice the change of direction. This CP-map
$c$ is called a \emph{channel} if it preserves the unit/identity
matrix: $c(\idmap) = \idmap$. It may be called \emph{subchannel} if
$c(\idmap) \leq \idmap$. Each CP-map $c\colon \mathcal{B}(\mathscr{K})
\rightarrow \mathcal{B}(\mathscr{H})$ has a `dagger', written as
$c^{\#}\colon \mathcal{B}(\mathscr{H}) \rightarrow
\mathcal{B}(\mathscr{K})$, so that $\HSip{c(A)}{B} =
\HSip{A}{c^{\#}(B)}$, that is, $\tr(c(A)^{\dag}B) =
\tr(A^{\dag}c^{\#}(B))$.

For a channel $c \colon \mathscr{H} \rightarrow \mathscr{K}$ and a
predicate (effect) $q$ on $\mathscr{K}$ we define predicate
transformation via function application $c \ll q \coloneqq c(q)$.
Similarly, for a state $\sigma$ on $\mathscr{H}$ we define state
transformation via the dagger of the channel, as: $c \gg \sigma
\coloneqq c^{\#}(\sigma)$. Then, using that positive operators are
self-adjoint, we get the same
relation~\eqref{eqn:classicalvaliditytransformation} between validity
and state/predicate transformation as in the classical case:
\begin{equation}
\label{eqn:quantumvaliditytransformation}
\mystretch
\begin{array}{rcl}
c \gg \sigma \models q
\hspace*{\arraycolsep}=\hspace*{\arraycolsep}
\tr\big(c^{\#}(s)\,q\big)
& = &
\tr\big(c^{\#}(s)\,q^{\dag}\big)
\\
& = &
\tr\big(s\,c(q)^{\dag}\big)
\hspace*{\arraycolsep}=\hspace*{\arraycolsep}
\tr\big(s\,c(q)\big)
\hspace*{\arraycolsep}=\hspace*{\arraycolsep}
s \models c \ll q.
\end{array}
\end{equation}

\begin{definition}
\label{def:assert}
Let $p$ be a (quantum) predicate on Hilbert space $\mathscr{H}$. It gives
rise to a subchannel $\asrt_{p} \colon \mathscr{H} \rightarrow \mathscr{H}$
defined by:
\[ \begin{array}{rcl}
\asrt_{p}(A)
& \coloneqq &
\sqrt{p}\,A\,\sqrt{p}.
\end{array} \]
\end{definition}

This assert map $\asrt_p$ plays a fundamental role in effectus theory,
see~\cite{Jacobs15d,ChoJWW15b}, for instance because it allows us to
define sequential conjunction~\eqref{AndthenEqn} via predicate
transformation as $p \andthen q = \asrt_{p} \ll q$.

\begin{rem}
\label{rem:daggerconditioning}
States/predicates on $\mathscr{H}$ are special instances of CP-maps
$\C \rightarrow \mathscr{H}$, resp.\ $\mathscr{H}\rightarrow\C$. If we
consider them as such channels, we can take their dagger
$(-)^{\#}$. Then we can relate upper and lower conditioning via an
exchange, namely as: $\sigma|^{p} = p^{\#}|_{\sigma^{\#}}$. This
re-formulation may be useful in a more general setting.
\end{rem}


\subsection{Representation of quantum channels}

As mentioned, a channel $c\colon \mathscr{H} \rightarrow \mathscr{K}$
is a (completely positive) linear function $\mathcal{B}(\mathscr{K})
\rightarrow \mathcal{B}(\mathscr{H})$ between spaces of operators.
Let's assume $\mathscr{H},\mathscr{K}$ have dimensions $n,m$,
respectively. The space of operators $\mathcal{B}(\mathscr{K})$ then
has dimension $m\times m$, so that the channel $c$ is determined by
its values on the $m\times m$ base vectors $\ket{i}\bra{j}$ of
$\mathcal{B}(\mathscr{K})$. Thus, the channel $c$ is determined by
$m\times m$ matrices of size $n\times n$, as in:
\begin{equation}
\label{diag:channelmatrix}
\begin{array}{cl}
\left(\begin{array}{ccc}
\left(\,\fbox{\strut\rule[-0.5em]{0em}{0em}$n\times n$}\,\right)
& \quad\cdots\quad &
\left(\,\fbox{\strut\rule[-0.5em]{0em}{0em}$n\times n$}\,\right)
\\
\vdots & & \vdots
\\
\left(\,\fbox{\strut\rule[-0.5em]{0em}{0em}$n\times n$}\,\right)
& \quad\cdots\quad &
\left(\,\fbox{\strut\rule[-0.5em]{0em}{0em}$n\times n$}\,\right)
\end{array}\right) & 
\raisebox{+0.7em}{$\underset{\underset{\textstyle\downarrow}{\textstyle m}}{\textstyle\uparrow}$}
\\
\leftarrow\!m\!\rightarrow 
\end{array}
\end{equation}

\noindent The matrix entries of the channel $c$ will be written via
double indexing, as $c_{k\ell,ij}$ for $1 \leq k,\ell \leq m$ and
$1\leq i,j \leq n$. 

This matrix representation of a quantum channel is used in EfProb. It
is convenient, for instance because parallel composition $\otimes$ of
channels can simply be done by Kronecker multiplication of their
(outer) matrices~\eqref{diag:channelmatrix}. We briefly describe how
predicate and state transformation works.

Let $q$ be a predicate on $\mathscr{K}$, represented as a $m\times m$
matrix. Predicate transformation $c \ll q$ is done simply by linear
extension. It yields an $n\times n$ matrix, forming a predicate on
$\mathscr{H}$, via:
\begin{equation}
\label{eqn:qpredtransform}
\begin{array}{rcl}
c \ll q
& \coloneqq &
\sum_{k,\ell}\, q_{k\ell}\cdot c_{k\ell}.
\end{array}
\end{equation}

\noindent In the other direction we do state transformation essentially
via the dagger $c^{\#}$ of the channel $c$. Explicitly, it works as
follows. Let $\sigma$ be a state of $\mathscr{H}$, represented by
a $n\times n$ matrix. Then we obtain the transformed state $c \gg \sigma$
as an $m\times m$ matrix given by computing traces:
\begin{equation}
\label{eqn:qstatetransform}
\begin{array}{rcl}
\big(c \gg \sigma\big)_{k\ell}
& \coloneqq &
\tr(c_{\ell k}\,\sigma).
\end{array}
\end{equation}

\noindent Notice the change of order of indices: at position
$(k,\ell)$ of $c \gg \sigma$ we use the inner matrix $c_{\ell k}$
from~\eqref{diag:channelmatrix}. The reason is the implicit use of the
Hilbert-Schmidt inner product, given by $\HSip{A}{B} =
\tr(A^{\dag}\cdot B)$, where the dagger involves a conjugate transpose.


\subsection{Quantum pairing and extraction}

The pairing of a classical state and a channel in~\eqref{diag:pairing}
involves a copier $\copier$. It does not exist in general in a quantum
setting because of the `no-cloning' theorem. But we do have `cup'
states $\cup$ with maximal entanglement. They are basis dependent:
given a finite-dimensional Hilbert space $\mathscr{H}$ with
orthonormal basis $\big(\ket{i}\big)$ of size $n$, we can for a state
$\cup$ of $\mathscr{H}\otimes\mathscr{H}$ as $\cup =
\frac{1}{n}\sum_{i,j}\ket{ii}\bra{jj}$. Similarly, there is `cap'
predicate $\cap$. The quantum pairing and extraction operations that
we describe in this subsection are due to~\cite{LeiferS13}. But the
more abstract description in terms of cups and caps does not occur
there.  These operations depend on a choice of basis.


\auxproof{
We first unravel the definition of $\cup$ and then illustrate that
it equalis the EfProb implementation, for $n=2$ and $n=3$.
\[ \begin{array}{rcl}
2\cup_{2}
& = &
\left(\begin{smallmatrix} 1 \\ 0 \\ 0 \\ 0 \end{smallmatrix}\right)
\left(\begin{smallmatrix} 1 & 0 & 0 & 0 \end{smallmatrix}\right)
+
\left(\begin{smallmatrix} 1 \\ 0 \\ 0 \\ 0 \end{smallmatrix}\right)
\left(\begin{smallmatrix} 0 & 0 & 0 & 1 \end{smallmatrix}\right)
+
\left(\begin{smallmatrix} 0 \\ 0 \\ 0 \\ 1 \end{smallmatrix}\right)
\left(\begin{smallmatrix} 1 & 0 & 0 & 0 \end{smallmatrix}\right)
+
\left(\begin{smallmatrix} 0 \\ 0 \\ 0 \\ 1 \end{smallmatrix}\right)
\left(\begin{smallmatrix} 0 & 0 & 0 & 1 \end{smallmatrix}\right)
\\
& = &
\left(\begin{smallmatrix}
1 & 0 & 0 & 1 \\ 
0 & 0 & 0 & 0 \\ 
0 & 0 & 0 & 0 \\ 
1 & 0 & 0 & 1
\end{smallmatrix}\right)
\\
& = &
\left(\begin{smallmatrix} 1 \\ 0 \\ 0 \\ 1 \end{smallmatrix}\right)
\left(\begin{smallmatrix} 1 & 0 & 0 & 1 \end{smallmatrix}\right)
\\
3\cup_{3}
& = &
\left(\begin{smallmatrix} 1 \\ 0 \\ 0 \\ 0 \\ 0 \\ 0 \\ 0 \\ 0 \\ 0 \end{smallmatrix}\right)
\left(\begin{smallmatrix} 1 & 0 & 0 & 0 & 0 & 0 & 0 & 0 & 0 \end{smallmatrix}\right)
+
\left(\begin{smallmatrix} 1 \\ 0 \\ 0 \\ 0 \\ 0 \\ 0 \\ 0 \\ 0 \\ 0 \end{smallmatrix}\right)
\left(\begin{smallmatrix} 0 & 0 & 0 & 0 & 1 & 0 & 0 & 0 & 0 \end{smallmatrix}\right)
+
\left(\begin{smallmatrix} 1 \\ 0 \\ 0 \\ 0 \\ 0 \\ 0 \\ 0 \\ 0 \\ 0 \end{smallmatrix}\right)
\left(\begin{smallmatrix} 0 & 0 & 0 & 0 & 0 & 0 & 0 & 0 & 1 \end{smallmatrix}\right)
\\
& = &
\left(\begin{smallmatrix} 0 \\ 0 \\ 0 \\ 0 \\ 1 \\ 0 \\ 0 \\ 0 \\ 0 \end{smallmatrix}\right)
\left(\begin{smallmatrix} 1 & 0 & 0 & 0 & 0 & 0 & 0 & 0 & 0 \end{smallmatrix}\right)
+
\left(\begin{smallmatrix} 0 \\ 0 \\ 0 \\ 0 \\ 1 \\ 0 \\ 0 \\ 0 \\ 0 \end{smallmatrix}\right)
\left(\begin{smallmatrix} 0 & 0 & 0 & 0 & 1 & 0 & 0 & 0 & 0 \end{smallmatrix}\right)
+
\left(\begin{smallmatrix} 0 \\ 0 \\ 0 \\ 0 \\ 1 \\ 0 \\ 0 \\ 0 \\ 0 \end{smallmatrix}\right)
\left(\begin{smallmatrix} 0 & 0 & 0 & 0 & 0 & 0 & 0 & 0 & 1 \end{smallmatrix}\right)
\\
& = &
\left(\begin{smallmatrix} 0 \\ 0 \\ 0 \\ 0 \\ 0 \\ 0 \\ 0 \\ 0 \\ 1 \end{smallmatrix}\right)
\left(\begin{smallmatrix} 1 & 0 & 0 & 0 & 0 & 0 & 0 & 0 & 0 \end{smallmatrix}\right)
+
\left(\begin{smallmatrix} 0 \\ 0 \\ 0 \\ 0 \\ 0 \\ 0 \\ 0 \\ 0 \\ 1 \end{smallmatrix}\right)
\left(\begin{smallmatrix} 0 & 0 & 0 & 0 & 1 & 0 & 0 & 0 & 0 \end{smallmatrix}\right)
+
\left(\begin{smallmatrix} 0 \\ 0 \\ 0 \\ 0 \\ 0 \\ 0 \\ 0 \\ 0 \\ 1 \end{smallmatrix}\right)
\left(\begin{smallmatrix} 0 & 0 & 0 & 0 & 0 & 0 & 0 & 0 & 1 \end{smallmatrix}\right)
\\
& = &
\left(\begin{smallmatrix} 
1 & 0 & 0 & 0 & 1 & 0 & 0 & 0 & 1 \\
0 & 0 & 0 & 0 & 0 & 0 & 0 & 0 & 0 \\
0 & 0 & 0 & 0 & 0 & 0 & 0 & 0 & 0 \\
0 & 0 & 0 & 0 & 0 & 0 & 0 & 0 & 0 \\
1 & 0 & 0 & 0 & 1 & 0 & 0 & 0 & 1 \\
0 & 0 & 0 & 0 & 0 & 0 & 0 & 0 & 0 \\
0 & 0 & 0 & 0 & 0 & 0 & 0 & 0 & 0 \\
0 & 0 & 0 & 0 & 0 & 0 & 0 & 0 & 0 \\
1 & 0 & 0 & 0 & 1 & 0 & 0 & 0 & 1 \\
\end{smallmatrix}\right)
\\
& = &
\left(\begin{smallmatrix} 1 \\ 0 \\ 0 \\ 0 \\ 1 \\ 0 \\ 0 \\ 0 \\ 1 \end{smallmatrix}\right)
\left(\begin{smallmatrix} 1 & 0 & 0 & 0 & 1 & 0 & 0 & 0 & 1 \end{smallmatrix}\right)
\end{array} \]
}

Given a state $\sigma$ of $\mathscr{H}$ and a channel $c\colon
\mathscr{H} \rightarrow \mathscr{K}$ we can thus form a joint state of
$\mathscr{H}\otimes\mathscr{K}$ via the `cup' state $\cup$ of
$\mathscr{H}\otimes\mathscr{H}$. Then we can define a pair state of
$\mathscr{H}\otimes\mathscr{K}$ via state transformation $\gg$ as in:
\begin{equation}
\label{eqn:quantumpairing}
\begin{array}{rclcrcl}
\pair(\sigma,c)
& \coloneqq &
\big(\asrt_{\sigma^{T}} \otimes c\big) \gg \cup
& \qquad\mbox{that is} \qquad
\bra{ik}\pair(\sigma,c)\ket{j\ell}
& = &
\overline{\big(\sqrt{\sigma}c_{k\ell}\sqrt{\sigma}\big)_{ij}}.
\end{array}
\end{equation} 

\noindent In the other direction, given a joint state $\tau$ of
$\mathscr{H}\otimes\mathscr{K}$ we write $\proj(\tau)$ for the transpose
of its first marginal, so:
\begin{equation}
\label{eqn:quantumproject}
\begin{array}{rclcrcl}
\proj(\tau)
& \coloneqq &
\margsign_{1}(\tau)^{T}
& \qquad\mbox{where}\qquad &
\bra{i}\margsign_{1}(\tau)\ket{j}
& \coloneqq &
\sum_{k} \bra{ik}\tau\ket{jk}.
\end{array}
\end{equation}

\noindent We extract a channel $\extract(\tau) \colon \mathscr{H}
\rightarrow \mathscr{K}$ from $\tau$ in the manner defined
in~\cite{LeiferS13}:
\begin{equation}
\label{eqn:quantumextract}
\begin{array}{rcl}
\extract(\tau)_{k\ell}
& \coloneqq &
\sum_{i,j} \overline{\bra{ik}\tau\ket{j\ell}} \cdot
   \big(\sqrt{\proj(\tau)^{-1}}\ket{i}\bra{j}\sqrt{\proj(\tau)^{-1}}\big).
\end{array}
\end{equation}

\noindent The next result is the analogue of
Lemma~\ref{lem:disintegration} about disintegration for classical
discrete probability.

\begin{proposition}[After~\cite{LeiferS13}]
\label{prop:quantumpairing}
A quandum state $\sigma$ and channel $c$, with matching types, can be
recovered from their pair, defined in~\eqref{eqn:quantumpairing}, via
projection~\eqref{eqn:quantumproject} and
extraction~\ref{eqn:quantumextract}
\[ \begin{array}{rclcrcl}
\proj(\pair(\sigma,c))
& = &
\sigma
& \hspace*{3em}\mbox{and}\hspace*{3em} &
\extract\big(\pair(\sigma,c)\big)
& = &
c.
\end{array} \]

\noindent Similarly, a joint state $\tau$ for which the transpose of
its first marginal $\proj(\tau)$, as defined above, is invertible can
be recovered as a pair, as on the left below. In addition, $\tau$'s
second marginal can be obtained via state transformation, as on the right:
\[ \begin{array}{rclcrcl}
\tau
& = &
\pair(\proj(\tau), \extract(\tau))
& \hspace*{7em} &
\margsign_{2}(\tau)
& = &
\extract(\tau) \gg \proj(\tau).
\end{array} \]
\end{proposition}

\begin{myproof}
We shall do the first equation and leave the others to the interested
reader.
\[ \mystretch\begin{array}{rcl}
\big(\proj(\pair(\sigma,c))\big)_{ij}
& = &
\sum_{k} \bra{jk}\pair(\sigma,c)\ket{ik}
\\
& \smash{\stackrel{\eqref{eqn:quantumpairing}}{=}} &
\sum_{k} \overline{\big(\sqrt{\sigma}c_{kk}\sqrt{\sigma}\big)_{ji}}
\\
& = &
\overline{\big(\sqrt{\sigma}\big(\sum_{k}c_{kk}\big)\sqrt{\sigma}\big)_{ji}}
\\
& = &
\overline{\sigma_{ji}} \qquad \mbox{since $c$ is unital, 
   \textit{i.e.}~$c \ll \one = \one$}
\\
& = &
\big(\sigma^{\dag}\big)_{ij}
\\
& = &
\sigma_{ij}.
\end{array} \]

\noindent The latter equation holds since a state is positive and thus
self-adjoint. \QED

\auxproof{
Similarly,
\[ \begin{array}{rcl}
\extract\big(\pair(\sigma,c)\big)_{k\ell}
& = &
\sum_{i,j} \overline{\bra{ik}\pair(\sigma,c)\ket{j\ell}} \cdot
   \big(\sqrt{\proj(\pair(\sigma,c))^{-1}}\ket{i}\bra{j}\sqrt{\proj(\pair(\sigma,c))^{-1}}\big)
\\
& = &
\sum_{i,j}  \big(\sqrt{\sigma}c_{k\ell}\sqrt{\sigma}\big)_{ij}
   \big(\sqrt{\sigma^{-1}}\ket{i}\bra{j}\sqrt{\sigma_{1}^{-1}}\big)
\\
& = &
\sum_{i,j,i',j'}  \sqrt{\sigma}_{ii'}c_{k\ell,i',j'}\sqrt{\sigma}_{jj'}
   \sqrt{\sigma^{-1}}\ket{i}\bra{j}\sqrt{\sigma_{1}^{-1}}
\\
& = &
\sum_{i',j'}  c_{k\ell,i',j'}
   \big(\sum_{i}  \sqrt{\sigma}_{ii'}\sqrt{\sigma^{-1}}\ket{i}\big)
   \big(\sum_{j}\bra{j}\sqrt{\sigma_{1}^{-1}}\sqrt{\sigma}_{jj'}\big)
\\
& = &
\sum_{i',j'}  c_{k\ell,i',j'} \ket{i'}\bra{j'}
\\
& = &
c_{k\ell}.
\end{array} \]

Next,
\[ \begin{array}{rcl}
\bra{ik}\pair(\proj(\tau), \extract(\tau))\ket{j\ell}
& = &
\overline{\big(\sqrt{\proj(\tau)}\extract(\tau)_{k\ell}\sqrt{\proj(\tau)}\big)_{ij}}
\\
& = &
\overline{\big(\sqrt{\proj(\tau)}\big(\sum_{i',j'} 
   \overline{\bra{i'k}\tau\ket{j'\ell}}
   \sqrt{\proj(\tau)^{-1}}\ket{i'}\bra{j'}\sqrt{\proj(\tau)^{-1}}\big)
   \sqrt{\proj(\tau)}\big)_{ij}}
\\
& = &
\big(\sum_{i',j'} \bra{i'k}\tau\ket{j'\ell} \ket{i'}\bra{j'}\big)_{ij}
\\
& = &
\bra{ik}\tau\ket{j\ell}.
\end{array} \]

And finally:
\[ \begin{array}{rcl}
\lefteqn{\big(\extract(\tau) \gg \proj(\tau)\big)_{k\ell}}
\\
& = &
\tr\big(\extract(\tau)_{\ell k} \, \proj(\tau)\big)
\\
& = &
\sum_{i} \big(\extract(\tau)_{\ell k} \, \proj(\tau)\big)_{ii}
\\
& = &
\sum_{i,j} \extract(\tau)_{\ell k,ij} \proj(\tau)_{ji}
\\
& = &
\sum_{i,j,i',i''}  \overline{\bra{i'\ell}\tau\ket{i''k}}
   \bra{i}\sqrt{\proj(\tau)^{-1}}\ket{i'}\bra{i''}\sqrt{\proj(\tau)^{-1}}\ket{j}
   \bra{j}\proj(\tau)\ket{i}
\\
& = &
\sum_{j,i',i''}  \overline{\bra{i'\ell}\tau\ket{i''k}}
   \bra{j}\sqrt{\proj(\tau)}\ket{i'}\bra{i''}\sqrt{\proj(\tau)^{-1}}\ket{j}
\\
& = &
\sum_{i',i''}  \overline{\bra{i'\ell}\tau\ket{i''k}}
   \ket{i'}\bra{i''}\idmap\ket{i'}
\\
& = &
\sum_{i} \overline{\bra{i\ell}\tau\ket{ik}}
\\
& = &
\sum_{i} \big(\tau^{\dag}\big)_{k\ell,ii}
\\
& = &
\sum_{i} \tau_{k\ell,ii}
\\
& = &
\big(\marg{\tau}{[0,1]}\big)_{k\ell}.
\end{array} \]
}
\end{myproof}

As an aside, for readers who are comfortable with diagrammatic
notation (see \textit{e.g.}~\cite{Selinger11,CoeckeK16}) one can write:
\vspace*{-0.8em}
\[ \pair(\sigma,c)
\;\;
=
\;\;
\vcenter{\hbox{%
\begin{tikzpicture}[font=\small]
\node[arrow box] (a) at (-1,0) {$\asrt_{\sigma^T}$};
\node[arrow box] (c) at (0,0) {$c$};
\draw (0.0,-0.3) to (c);
\draw (0,-0.3) arc(0.0:-180:0.5);
\draw (-1,-0.3) to (a);
\draw (a) to (-1,0.5);
\draw (c) to (0,0.5);
\end{tikzpicture}}}
\hspace*{3em}
\proj(\tau)
\;\;
=
\;\;
\vcenter{\hbox{%
\begin{tikzpicture}[font=\small]
\node[state] (t) at (0,0) {$\;\;\tau\;\;$};
\coordinate (t1) at ([xshiftu=-0.25]t);
\coordinate (t2) at ([xshiftu=0.25]t);
\node[discarder] (d) at ([yshiftu=0.3]t2) {};
\draw (t1) to ([yshiftu=0.2]t1);
\draw (t2) to (d);
\draw (-0.25,0.2) arc(0.0:180:0.4);
\draw (-1.05,-0.5) to (-1.05,0.2);
\end{tikzpicture}}}
\hspace*{3em}
\extract(t)
\;\;
=
\;\;
\vcenter{\hbox{%
\begin{tikzpicture}[font=\small]
\node[state] (t) at (0,0) {$\;\;\tau\;\;$};
\coordinate (t1) at ([xshiftu=-0.25]t);
\coordinate (t2) at ([xshiftu=0.25]t);
\node[arrow box, minimum height=1.5em] (a) at (-2.05,-0.3) %
   {\raisebox{0.2em}{$\asrt_{\proj(\tau)^{-1}}$}};
\draw (t1) to ([yshiftu=0.1]t1);
\draw (t2) to ([yshiftu=0.7]t2);
\draw (-0.25,0.1) arc(0.0:180:0.9);
\draw (a) to ([yshiftu=+0.5]a);
\draw (a) to ([yshiftu=-0.6]a);
\end{tikzpicture}}}
\]

\section{A quantum Bayesian Inference Theorem}\label{sec:quantuminference}

This section contains the main result of this paper, namely the
quantum analogue of Theorem~\ref{thm:classicalconditioning}. It
describes how conditioning of a joint state can also be performed via
the extracted channel. The novelty in our quantum description is that
we need both lower and upper conditioning to capture what is going on.

\begin{theorem}
\label{thm:quantuminference}
Let $\tau$ be a state of $\mathscr{H}\otimes\mathscr{K}$ and let $p,q$
be predicates, on $\mathscr{H}$ and on $\mathscr{K}$ respectively.
Then:
\[ \begin{array}[b]{rclcrcl}
\margsign_{2}\big(\tau|_{p \otimes \one}\big)
& = &
\extract(\tau) \gg (\proj(\tau)|^{p^{T}})
& \qquad\mbox{and}\qquad &
\margsign_{1}\big(\tau|_{\one \otimes q}\big)
& = &
\big(\proj(\tau)|^{\extract(\tau) \ll q}\big)^{T}.
\end{array} \eqno{\QEDbox} \]
\end{theorem}

The proof is ommitted since it involves rather long and boring matrix
calculations. Instead we include a random test: the quantum versions
of pairing / projection / extraction and lower / upper conditioning
have been implemented in EfProb. They can be used to test
Theorem~\ref{thm:quantuminference} as below, by generating an
arbitrary state \pythoninline{t}, in this case of type $\C^{3}\otimes
\C^{5}$, together with arbitrary (suitably typed) predicates.  The
EfProb notation for lower and upper conditioning is \pythoninline{/}
and \pythoninline{^}.
\begin{python}
>>> t = random_state([3,5])
>>> t1 = (t % [1,0]).transpose()
>>> e = extract(t)
>>> p = random_pred([3])
>>> q = random_pred([5])
>>> t / (p @ truth([5])) % [0,1] == e >> (t1 ^ p.transpose())
True
>>> t / (truth([3]) @ q) % [1,0] == (t1 ^ (e << q)).transpose()
True
\end{python}

\noindent The two equality tests \pythoninline{==} involve $5\times 5$
and $3\times 3$ matrices of complex numbers.

In the equations in Theorem~\ref{thm:quantuminference} we perform
lower conditioning on the joint state. One may ask if there are also
`dual' equations where upper conditioning on the joint state is
re-described via state/predicate transformation. We have not found
them.

\subsection*{Acknowledgements}

Thanks to Kenta Cho and Alex Kissinger for helpful feedback and
discussions.

\bibliographystyle{plain}

\end{document}

%% file: tikzdefs.tex
\newlength\stateheight
\newlength\minimumstatewidth
\tikzset{width/.initial=\minimummorphismwidth}
\tikzset{colour/.initial=white}

\newif\ifblack\pgfkeys{/tikz/black/.is if=black}
\newif\ifwedge\pgfkeys{/tikz/wedge/.is if=wedge}
\newif\ifvflip\pgfkeys{/tikz/vflip/.is if=vflip}
\newif\ifhflip\pgfkeys{/tikz/hflip/.is if=hflip}
\newif\ifhvflip\pgfkeys{/tikz/hvflip/.is if=hvflip}

\pgfdeclarelayer{foreground}
\pgfdeclarelayer{background}
\pgfsetlayers{background,main,foreground}

\def\thickness{0.4pt}

\makeatletter
\pgfkeys{%
  /tikz/on layer/.code={
    \pgfonlayer{#1}\begingroup
    \aftergroup\endpgfonlayer
    \aftergroup\endgroup
  },
  /tikz/node on layer/.code={
    \gdef\node@@on@layer{%
      \setbox\tikz@tempbox=\hbox\bgroup\pgfonlayer{#1}\unhbox\tikz@tempbox\endpgfonlayer\pgfsetlinewidth{\thickness}\egroup}
    \aftergroup\node@on@layer
  },
  /tikz/end node on layer/.code={
    \endpgfonlayer\endgroup\endgroup
  }
}
\def\node@on@layer{\aftergroup\node@@on@layer}
\makeatother

\makeatletter
\pgfdeclareshape{state}
{
    \savedanchor\centerpoint
    {
        \pgf@x=0pt
        \pgf@y=0pt
    }
    \anchor{center}{\centerpoint}
    \anchorborder{\centerpoint}
    \saveddimen\overallwidth
    {
        \pgf@x=3\wd\pgfnodeparttextbox
        \ifdim\pgf@x<\minimumstatewidth
            \pgf@x=\minimumstatewidth
        \fi
    }
    \savedanchor{\upperrightcorner}
    {
        \pgf@x=.5\wd\pgfnodeparttextbox
        \pgf@y=.5\ht\pgfnodeparttextbox
        \advance\pgf@y by -.5\dp\pgfnodeparttextbox
    }
    \anchor{A}
    {
        \pgf@x=-\overallwidth
        \divide\pgf@x by 4
        \pgf@y=0pt
    }
    \anchor{B}
    {
        \pgf@x=\overallwidth
        \divide\pgf@x by 4
        \pgf@y=0pt
    }
    \anchor{text}
    {
        \upperrightcorner
        \pgf@x=-\pgf@x
        \ifhflip
            \pgf@y=-\pgf@y
            \advance\pgf@y by 0.4\stateheight
        \else
            \pgf@y=-\pgf@y
            \advance\pgf@y by -0.4\stateheight
        \fi
    }
    \backgroundpath
    {
       \begin{pgfonlayer}{foreground}
        \pgfsetstrokecolor{black}
        \pgfsetlinewidth{\thickness}
        \pgfpathmoveto{\pgfpoint{-0.5*\overallwidth}{0}}
        \pgfpathlineto{\pgfpoint{0.5*\overallwidth}{0}}
        \ifhflip
            \pgfpathlineto{\pgfpoint{0}{\stateheight}}
        \else
            \pgfpathlineto{\pgfpoint{0}{-\stateheight}}
        \fi
        \pgfpathclose
        \ifblack
            \pgfsetfillcolor{black!50}
            \pgfusepath{fill,stroke}
        \else
            \pgfusepath{stroke}
        \fi
       \end{pgfonlayer}
    }
}

\pgfdeclareshape{my ground}
{
  \inheritsavedanchors[from=rectangle ee]
  \inheritanchor[from=rectangle ee]{center}
  \inheritanchor[from=rectangle ee]{north}
  \inheritanchor[from=rectangle ee]{south}
  \inheritanchor[from=rectangle ee]{east}
  \inheritanchor[from=rectangle ee]{west}
  \inheritanchor[from=rectangle ee]{north east}
  \inheritanchor[from=rectangle ee]{north west}
  \inheritanchor[from=rectangle ee]{south east}
  \inheritanchor[from=rectangle ee]{south west}
  \inheritanchor[from=rectangle ee]{input}
  \inheritanchor[from=rectangle ee]{output}
  \anchorborder{\csname pgf@anchor@my ground@input\endcsname}

  \backgroundpath{
    \pgf@process{\pgfpointadd{\southwest}{\pgfpoint{\pgfkeysvalueof{/pgf/outer xsep}}{\pgfkeysvalueof{/pgf/outer ysep}}}}
    \pgf@xa=\pgf@x \pgf@ya=\pgf@y
    \pgf@process{\pgfpointadd{\northeast}{\pgfpointscale{-1}{\pgfpoint{\pgfkeysvalueof{/pgf/outer xsep}}{\pgfkeysvalueof{/pgf/outer ysep}}}}}
    \pgf@xb=\pgf@x \pgf@yb=\pgf@y
    \pgfpathmoveto{\pgfqpoint{\pgf@xa}{\pgf@ya}}
    \pgfpathlineto{\pgfqpoint{\pgf@xa}{\pgf@yb}}
    \pgfmathsetlength\pgf@xa{.5\pgf@xa+.5\pgf@xb}
    \pgfmathsetlength\pgf@yc{.16666\pgf@yb-.16666\pgf@ya}
    \advance\pgf@ya by\pgf@yc
    \advance\pgf@yb by-\pgf@yc
    \pgfpathmoveto{\pgfqpoint{\pgf@xa}{\pgf@ya}}
    \pgfpathlineto{\pgfqpoint{\pgf@xa}{\pgf@yb}}
    \advance\pgf@ya by\pgf@yc
    \advance\pgf@yb by-\pgf@yc
    \pgfpathmoveto{\pgfqpoint{\pgf@xb}{\pgf@ya}}
    \pgfpathlineto{\pgfqpoint{\pgf@xb}{\pgf@yb}}
  }
}
\makeatother

\tikzset{inline text/.style =
  {text height=1.2ex, text depth=0.25ex,yshift=0.5mm}}
\tikzset{arrow box/.style =
  {rectangle,inline text,fill=white,draw,
    minimum height=5mm,yshift=-0.5mm,minimum width=5mm}}
\tikzset{dot/.style =
  {inner sep=0mm,minimum width=1mm,minimum height=1mm,
    draw,shape=circle}}
\tikzset{white dot/.style = {dot,fill=white,text depth=-0.2mm}}
\tikzset{scalar/.style = {diamond,draw,inner sep=1pt}}

\tikzset{copier/.style = {dot,fill,text depth=-0.2mm}}
\tikzset{discarder/.style = {my ground,draw,inner sep=0pt,
    minimum width=4.2pt,minimum height=11.2pt,anchor=input,rotate=90}}

\tikzset{uniform/.style = {my ground,draw,inner sep=0pt,
    minimum width=4.2pt,minimum height=11.2pt,anchor=input,rotate=270}}

\tikzset{xshiftu/.style = {shift = {(#1, 0)}}}
\tikzset{yshiftu/.style = {shift = {(0, #1)}}}

%% file: lower_upper_conditioning.bbl
\begin{thebibliography}{10}
\providecommand{\url}[1]{\texttt{#1}}
\providecommand{\href}[2]{\texttt{#2}}
\providecommand{\urlalt}[2]{\href{#1}{#2}}
\providecommand{\doi}[1]{doi:\urlalt{http://dx.doi.org/#1}{#1}}
\providecommand{\arxiv}[1]{\urlalt{http://arxiv.org/abs/#1}{#1}}

\bibitem{AllenBHLS17}
J.-M. Allen, J.~Barrett, D.~Horsman, C.~Lee, and R.~Spekkens.
\newblock Quantum common causes and quantum causal models.
\newblock {\em Phys. Rev. X}, 7(3):031021, 2017.
\newblock \doi{10.1103/PhysRevX.7.031021}.

\bibitem{Barber12}
D.~Barber.
\newblock {\em Bayesian Reasoning and Machine Learning}.
\newblock Cambridge Univ. Press, 2012.
\newblock \doi{10.1017/CBO9780511804779}.
\newblock Publicly available via
  \url{http://web4.cs.ucl.ac.uk/staff/D.Barber/pmwiki/pmwiki.php?n=Brml.HomePage}.
  
\bibitem{BernardoS00}
J.~Bernardo and A.~Smith.
\newblock {\em Bayesian Theory}.
\newblock John Wiley \& Sons, 2000.


\bibitem{ChoJ17a}
K.~Cho and B.~Jacobs.
\newblock Disintegration and {Bayesian} inversion, both abstractly and
  concretely.
\newblock See arXiv:\arxiv{1709.00322}, 2017.

\bibitem{ChoJ17b}
K.~Cho and B.~Jacobs.
\newblock The {EfProb} library for probabilistic calculations.
\newblock In F.~Bonchi and B.~K{\"o}nig, editors, {\em Conference on Algebra
  and Coalgebra in Computer Science (CALCO 2017)}, volume~72 of {\em LIPIcs}.
  Schloss Dagstuhl, 2017.
\newblock \doi{10.4230/LIPIcs.CALCO.2017.25}.

\bibitem{ChoJWW15b}
K.~Cho, B.~Jacobs, A.~Westerbaan, and B.~Westerbaan.
\newblock An introduction to effectus theory.
\newblock See arXiv:\arxiv{1512.05813}, 2015.

\bibitem{CoeckeK16}
B.~Coecke and A.~Kissinger.
\newblock {\em Picturing Quantum Processes. A First Course in Quantum Theory
  and Diagrammatic Reasoning}.
\newblock Cambridge Univ. Press, 2016.
\newblock \doi{10.1017/9781316219317}.

\bibitem{CoeckeS12}
B.~Coecke and R.~Spekkens.
\newblock Picturing classical and quantum {Bayesian} inference.
\newblock {\em Synthese}, 186(3):651--696, 2012.
\newblock \doi{10.1007/s11229-011-9917-5}

\bibitem{Fong12}
B.~Fong.
\newblock Causal theories: A categorical perspective on {Bayesian} networks.
\newblock Master's thesis, Univ.\ of Oxford, 2012.
\newblock See arXiv:\arxiv{1301.6201}.


\bibitem{GudderG02}
S.~Gudder and R.~Greechie.
\newblock Sequential products on effect algebras.
\newblock {\em Reports on Math. Physics}, 49(1):87--111, 2002.
\newblock \doi{10.1016/S0034-4877(02)80007-6}.

\bibitem{Hayashi85}
S.~Hayashi.
\newblock Adjunction of semifunctors: categorical structures in nonextensional
  lambda calculus.
\newblock {\em Theor. Comp. Sci.}, 41:95--104, 1985.
\newblock \doi{10.1016/0304-3975(85)90062-3}.

\bibitem{HoofmanM95}
R.~Hoofman and I.~Moerdijk.
\newblock A remark on the theory of semi-functors.
\newblock {\em Math. Struct. in Comp. Sci.}, 5(1):1--8, 1995.
\newblock \doi{10.1017/S096012950000061X}.

\bibitem{Jacobs15d}
B.~Jacobs.
\newblock New directions in categorical logic, for classical, probabilistic and
  quantum logic.
\newblock {\em Logical Methods in Comp. Sci.}, 11(3), 2015.
\newblock \doi{10.2168/LMCS-11(3:24)2015}.

\bibitem{Jacobs17a}
B.~Jacobs.
\newblock From probability monads to commutative effectuses.
\newblock {\em Journ. of Logical and Algebraic Methods in Programming},
  94:200--237, 2017.
\newblock \doi{10.1016/j.jlamp.2016.11.006}
  
\bibitem{Jacobs17f}
B.~Jacobs.
\newblock Quantum effect logic in cognition.
\newblock {\em Journ. Math. Psychology}, 81:1--10, 2017.
\newblock \doi{10.1016/j.jmp.2017.08.004}.

\bibitem{JacobsZ16}
B.~Jacobs and F.~Zanasi.
\newblock A predicate/state transformer semantics for {Bayesian} learning.
\newblock In L.~Birkedal, editor, {\em Math. Found. of Programming Semantics},
  number 325 in Elect. Notes in Theor. Comp. Sci., pages 185--200. Elsevier,
  Amsterdam, 2016.
\newblock \doi{10.1016/j.entcs.2016.09.038}.
  
\bibitem{JacobsZ17}
B.~Jacobs and F.~Zanasi.
\newblock A formal semantics of influence in {Bayesian} reasoning.
\newblock In K.~Larsen, H.~Bodlaender, and J.-F. Raskin, editors, {\em Math.
  Found. of Computer Science}, volume~83 of {\em LIPIcs}, pages 21:1--21:14.
  Schloss Dagstuhl, 2017.
\newblock \doi{10.4230/LIPIcs.MFCS.2017.21}.
  
\bibitem{JacobsZ18}
B.~Jacobs and F.~Zanasi.
\newblock The logical essentials of {Bayesian} reasoning.
\newblock See arXiv:\arxiv{1804.01193}, 2018.

\bibitem{KollerF09}
D.~Koller and N.~Friedman.
\newblock {\em Probabilistic Graphical Models. Principles and Techniques}.
\newblock {MIT} Press, Cambridge, MA, 2009.

\bibitem{LeiferS13}
M.~Leifer and R.~Spekkens.
\newblock Towards a formulation of quantum theory as a causally neutral theory
  of {Bayesian} inference.
\newblock {\em Phys. Rev. A}, 88(5):052130, 2013.
\newblock \doi{10.1103/PhysRevA.88.052130}.

\bibitem{LeiferS14}
M.~Leifer and R.~Spekkens.
\newblock A {Bayesian} approach to compatibility, improvement, and pooling of
  quantum states.
\newblock {\em Journ. of Physics A: Mathematical and Theoretical},
  47(27):275301, 2014.
\newblock \doi{10.1088/1751-8113/47/27/275301}.
  
\bibitem{NielsenC00}
M.~Nielsen and I.~Chuang.
\newblock {\em Quantum Computation and Quantum Information}.
\newblock Cambridge Univ. Press, 2000.
\newblock \doi{10.1017/CBO9780511976667}.

\bibitem{Pearl88}
J.~Pearl.
\newblock {\em Probabilistic Reasoning in Intelligent Systems: Networks of
  Plausible Inference}.
\newblock Graduate Texts in Mathematics 118. Morgan Kaufmann, 1988.

\bibitem{PienaarB15}
J.~Pienaar and {\v{C}}.~Brukner.
\newblock A graph-separation theorem for quantum causal models.
\newblock {\em New Journ. of Physics}, 17:073020, 2015.
\newblock \doi{10.1088/1367-2630/17/7/073020}.

\bibitem{Selinger11}
P.~Selinger.
\newblock A survey of graphical languages for monoidal categories.
\newblock In B.~Coecke, editor, {\em New Structures in Physics}, number 813 in
  Lect. Notes Physics, pages 289--355. Springer, Berlin, 2011.
\newblock \doi{10.1007/978-3-642-12821-9\_4}
  
\end{thebibliography}
